\documentclass[apj]{emulateapj}

\usepackage{graphicx}

\usepackage{natbib}
\usepackage[usenames]{color}

\newcommand{\bd}{\begin{displaymath}}
\newcommand{\ed}{\end{displaymath}}
\newcommand{\be}{\begin{equation}}
\newcommand{\ee}{\end{equation}}
\newcommand{\beaa}{\begin{eqnarray*}}
\newcommand{\eeaa}{\end{eqnarray*}}
\newcommand{\bea}{\begin{eqnarray}}
\newcommand{\eea}{\end{eqnarray}}

\newcommand\anguita{COSMOS\,5921$+$0638}

\def\hst{\textit{HST}}

\def\zl{z_{\rm l}}
\def\zs{z_{\rm s}}

\def\rein{r_{\rm ein}}

\def\zmag{m_{\rm z}}
\def\chith{\chi^2_{\text{th}}}
\def\Chitah{{\sc Chitah}}

\newcommand{\sref}[1]{Section~\ref{#1}}
\newcommand{\fref}[1]{Figure~\ref{#1}}
\newcommand{\fsref}[1]{Figures~\ref{#1}}
\newcommand{\tref}[1]{Table~\ref{#1}}
\newcommand{\eref}[1]{Equation~(\ref{#1})}

\def\ie{{i.e.,}}

\begin{document}

\title{\Chitah: Strong-gravitational-lens hunter in imaging surveys}

\author{James H.~H.~Chan\altaffilmark{1,2},
Sherry H.~Suyu\altaffilmark{2},
Tzihong Chiueh\altaffilmark{1,3,4},
Anupreeta More\altaffilmark{5},
Philip J.~Marshall\altaffilmark{6},
Jean Coupon\altaffilmark{7}, 
Masamune Oguri\altaffilmark{8,9,10},
Paul Price\altaffilmark{11}
}
\altaffiltext{1}{Department of Physics, National Taiwan University, 10617 Taipei, Taiwan}
\altaffiltext{2}{Institute of Astronomy and Astrophysics, Academia Sinica, P.O.~Box 23-141, Taipei 10617, Taiwan}
\altaffiltext{3}{Institute of Astrophysics, National Taiwan University, 10617 Taipei, Taiwan}
\altaffiltext{4}{Center for Theoretical Sciences, National Taiwan University, 10617 Taipei, Taiwan}
\altaffiltext{5}{Kavli IPMU (WPI), UTIAS, The University of Tokyo, Kashiwa, Chiba 277-8583, Japan }
\altaffiltext{6}{Kavli Institute for Particle Astrophysics and Cosmology, Stanford University, 452 Lomita Mall, Stanford, CA 94035, USA} 
\altaffiltext{7}{Astronomical Observatory of the University of Geneva, ch. d'Ecogia 16, 1290 Versoix, Switzerland}
\altaffiltext{8}{Kavli Institute for the Physics and Mathematics of the Universe (Kavli IPMU, WPI),
 University of Tokyo, 5-1-5 Kashiwanoha, Kashiwa-shi, Chiba 277-8583, Japan}
\altaffiltext{9}{Research Center for the Early Universe, University of Tokyo, 7-3-1 Hongo, Bunkyo-ku, Tokyo 113-0033, Japan}
\altaffiltext{10}{Department of Physics, University of Tokyo, 7-3-1 Hongo, Bunkyo-ku, Tokyo 113-0033, Japan}
\altaffiltext{11}{Department of Astrophysical Sciences, Princeton University, Princeton, NJ 08544, USA}

\email{d00222002@ntu.edu.tw}
\shorttitle{\Chitah: hunter of strong lenses}
\shortauthors{Chan, Suyu, Chiueh et al.}


\begin{abstract}

Strong gravitationally lensed quasars provide powerful means to study
galaxy evolution and cosmology.  Current and upcoming imaging surveys
will contain thousands of new lensed quasars, augmenting the existing
sample by at least two orders of magnitude.  To find such lens
systems, we built a robot, \Chitah, that hunts for lensed quasars
by modeling the configuration of the multiple quasar images.
Specifically, given an image of an object that might be a lensed
quasar, \Chitah\ first disentangles the light from the supposed
lens galaxy and the light from the multiple quasar images based on
color information.  A simple rule is designed to categorize the given
object as a potential four-image (quad) or two-image (double) lensed quasar
system.  The configuration of the identified quasar images is
subsequently modeled to classify whether the object is a lensed quasar
system.  We test the performance of \Chitah\ using simulated lens
systems based on the Canada-France-Hawaii Telescope Legacy Survey.
For bright quads with large image separations (with Einstein radius
$\rein>1\farcs1$) simulated using Gaussian point-spread functions,  
a high true-positive rate (TPR) of $\sim/$90\% and a low false-positive rate of $\sim$$3\%$
show that this is a promising approach to search for new lens systems.
We obtain high TPR for lens systems with $\rein\gtrsim0.5''$, so the
performance of \Chitah\ is set by the seeing.
We further feed a known gravitational lens system, \anguita, to \Chitah, and demonstrate that \Chitah\ is able to classify 
this real gravitational lens system successfully.  Our newly built
\Chitah\ is omnivorous and can hunt in any ground-based imaging
surveys.

\end{abstract}

\keywords{(galaxies:) quasars: individual (\anguita) --- gravitational lensing: strong --- methods: data analysis  }


\section{Introduction} 
\label{sec:intro}

Strong gravitational lensing occurs when light emitted from a source is deflected by a foreground lens object, resulting in multiple images.
Although lens systems are quite rare, we can use them to measure the mass distribution of foreground objects, from galaxies to galaxy clusters.
Moreover, the signal from background source objects is magnified so we can make use of this information to probe the high-redshift universe.

The first strong gravitational lens system, Q0957+561, was discovered by \citet{WalshEtal79}. 
This two-image lensed object provided the first opportunity to study cosmology through strong lensing tools. 
Since then, there have been many searches through imaging or spectroscopic surveys for lenses. 
However, most of them are aimed at detecting lensed galaxies rather than lensed quasars, including
 the Sloan Lens ACS Survey \citep[SLACS; e.g.,][]{BoltonEtal06SLACS}, the
 CFHTLS\footnote{Canada-France-Hawaii Telescope Legacy Survey.  See http://www.cfht.hawaii.edu/Science/CFHLS/ and links therein for a comprehensive description.} 
Strong Lensing Legacy Survey \citep[SL2S; e.g.,][]{CabanacEtal07, GavazziEtal12,MoreEtal12}, 
the BOSS Emission-line Lens Survey \cite[BELLS; e.g.,][]{BrownsteinEtal12}, 
the \hst Archive Galaxy-scale Gravitational Lens Search \citep[HAGGLeS;][]{MarshallEtal09}, 
\textit{Herschel} ATLAS \citep[H-ATLAS; e.g.,][]{NegrelloEtal10}, and the South Pole Telescope \citep[SPT; e.g.,][]{VieiraEtal13}.
Through these surveys, there are now a couple of hundred of strong lenses with different source populations.  
We expect that bigger samples will be discovered in ongoing imaging surveys \citep{OguriMarshall10}, 
such as the Hyper-Suprime-Cam (HSC) Survey \citep{MiyazakiEtal12}
 and the Dark Energy Survey (DES) \citep{SanchezEtail10}. 

Lensed quasars, although rarer than lensed galaxies, provide powerful means to study both galaxy evolution and cosmology.
For galaxy evolution,  
we can study galaxy mass structures and substructures through the use of the positions, shapes, and fluxes of lensed images \citep[e.g.,][]{SuyuEtal12, DalalKochanek02, VegettiEtal12}.
For cosmology, measuring time delays between multiple images allows us to determine the time-delay distance, 
which is sensitive to the Hubble constant, $H_0$ \citep[e.g.,][]{Refsdal64, CourbinEtal11, TewesEtal13, SuyuEtal10, SuyuEtal13}.
This quantity is one of the crucial cosmological parameters that sets
the age, size, and critical density of the universe.  By combining the
time delays with the stellar velocity dispersion of the lens,
we can also measure the angular diameter distance to the lens for
cosmological studies \citep[e.g.,][]{ParaficzHjorth09, JeeEtal14}.

Since lensed quasars are very useful, there have been several
undertakings to look for them with various suverys.  The Cosmic Lens
All-sky Survey \citep[CLASS;][]{MyersEtal03} discovered the largest
statistical sample of radio-loud gravitational lenses by obtaining
high-resolution images of flat-spectrum radio sources and identifying
the ones that showed multiple images. 
In the optical, the SDSS Quasar Lens Search
\citep[SQLS; e.g.,][]{OguriEtal06, OguriEtal08, OguriEtal12, InadaEtal08, InadaEtal10, InadaEtal12} has obtained the largest
lensed quasar sample to date based on both morphological and color
selection of spectroscopically confirmed quasars. 
\cite{JacksonEtal12} further combined the quasar samples from the SDSS and the UKIRT Infrared Deep Sky Survey (UKIDSS) to find small-separation or high-flux-ratio
lenses.
Another systematic approach has been proposed by \cite{KochanekEtal06} where all extended variable sources are identified as potential lenses.
Recently, \cite{AgnelloEtal15} proposed a novel way to select lens candidates through machine-learning algorithms.

We focus on an independent and effective way to detect lens systems automatically via modeling the quasar 
image configurations, as first demonstrated by \citet{MarshallEtal09}
who detected lenses in the \textit{Hubble Space Telescope} (\hst) archival
images via lens modeling as part of HAGGLeS.  The philosophy of
HAGGLeS is that for a lens candidate to be considered as such, its
imaging data must be able to be explained by a lens model.  
Therefore, they 
use a Singular Isothermal Sphere (SIS) as lens mass profile plus
external shear to fit the observed images of candidate lens objects.
However, the HAGGLeS robot aims at detecting lensed galaxies rather than lensed quasars. 
Inspired by HAGGLeS, we build a robot, \Chitah, to search for 
lensed quasars in imaging surveys via modeling.

\Chitah\ is an acronym for Chung-li He In-hsiang Tan Ao Hao, which is a direct transliteration from 
\includegraphics[height=10pt]{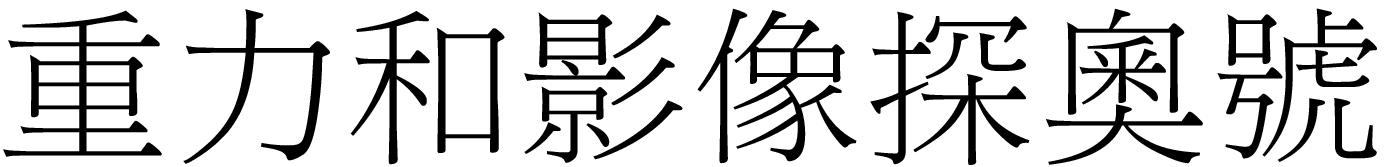}
that means a robot for explorations of gravitational imaging.
This robot is able to measure the positions of the lens galaxy and the multiple quasar images. 
We also employ a Singular Isothermal Ellipsoid (SIE) and a SIS as lens mass profiles 
to identify lenses with 
four-image and two-image configurations of quasar images (also known as ``quads'' and ``doubles''), respectively.

We design \Chitah\ with
multi-filter, high-resolution and signal-to-noise imaging data in
mind, \ie\ the HSC Survey and the Large Synoptic Survey Telescope (LSST). The
separation of the lens galaxy and quasar components for the modeling
will depend on data quality, and other approaches may be better suited
to poorer quality imaging data (e.g., LensTractor; P. J. Marshall et al. 2015, in
preparation).

This paper is organized as follows. 
In \sref{sec:procedure}, we detail the procedure of how \Chitah\
classifies lens candidates.  
We describe the simulated lenses based on CFHTLS data for educating
\Chitah\ in \sref{sec:simulation}, and present the results of the  
training in \sref{sec:results}.  We demonstrate that \Chitah\ can
successfully identify a real gravitational lens in
\sref{sec:application}.  We conclude in \sref{sec:summary}.
Magnitudes quoted in this paper are in AB magnitudes.


\section{\Chitah: Lens Finding Robot}
\label{sec:procedure}

\begin{table*}[t]
\caption{Lens and Quasar Colors and Brightnesses} 
\label{tab:case}
\begin{center}
\begin{tabular}{lllll}
\hline
 & Case 1 (Typical)  & Case 2 & Case 3 & Case 4\\ \hline
relative color and & quasar is bluer         & quasar is bluer & quasar is redder  & quasar is redder \\
 brightness & lens is brighter in z & quasar is brighter in z & lens is brighter in z & quasar is brighter in z \\
\hline\hline
$D_1 = g - \alpha z  $ \ \  (Equation (\ref{equ:qmap})) & $D_1 = \text{\rm quasar images}$ & $-D_1 = \text{\rm lens galaxy  }$ & $-D_1 = \text{\rm quasar images}$ & $D_1 = \text{\rm lens galaxy  }$ \\ 
$D_2 = z - \beta  D_1$ (Equation (\ref{lmap})) & $D_2 = \text{\rm lens galaxy  }$ & $ D_2 = \text{\rm quasar images}$ & $ D_2 = \text{\rm lens galaxy  }$ & $D_2 = \text{\rm quasar images}$ \\ 
\hline
\end{tabular}
\end{center}
Notes.  Columns 2-5 are the four possible scenarios for the colors and
brightnesses of the quasars and lens galaxies. 
We use the image cutouts in g-band and z-band labelled as $g$ and $z$ respectively.  In column 1, 
$\alpha$ is a scaling factor which scales the brightest pixel value in
$z$ to be the same as the corresponding pixel in $g$.  Similarly, 
$\beta$ is another scaling factor which scales the brightest pixel
value in $|D_1|$ to be the same as the corresponding pixel in $z$.   
For different cases, $D_1$ and $D_2$ yield either the lens light or the multiple images.  See \sref{subsec:separation} for details.
\end{table*}

The criterion for selecting a lensed system is based on the 
configuration of the quasar images. 
Therefore, we have to separate lens galaxy and quasar images, and then
identify the quasar image positions.
To separate lens and quasar images, we can make use of their color
information.  For simplicity, we use two imaging bands for constructing
the color.  We illustrate the method with g and z bands (which are
frequently available from large-scale imaging surveys), but the
method can be applied to any other two bands, provided they
are sufficiently separated in wavelength to distinguish the different
colors of the lens galaxies and quasars. 

\begin{figure}
\centering
\includegraphics[scale=0.4]{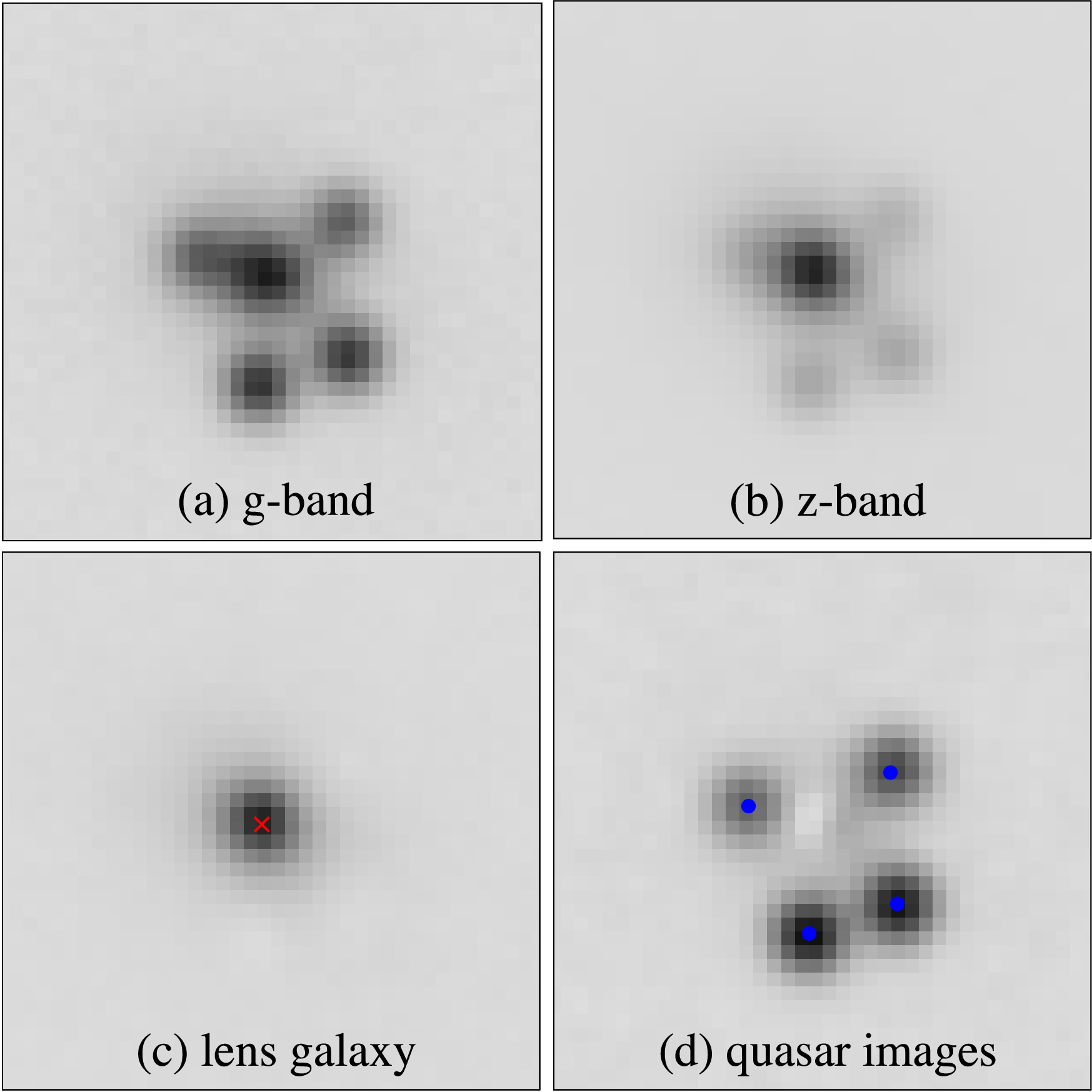}
\caption{An example of a simulated quad system. Panels (a) and (b):
  g-band and z-band cutouts, respectively.  
Panels (c) and (d): the lens galaxy and the quasar images,
respectively, which are separated based on color information and the
procedure described in \sref{subsec:separation}.  
The red cross in (c) is the estimated centroid of the lens light.
We identify the locations of quasar images, which we indicate with the
four blue dots in (d).}
\label{fig:quad_ex}
\end{figure}

\begin{figure}
\centering
\includegraphics[scale=0.4]{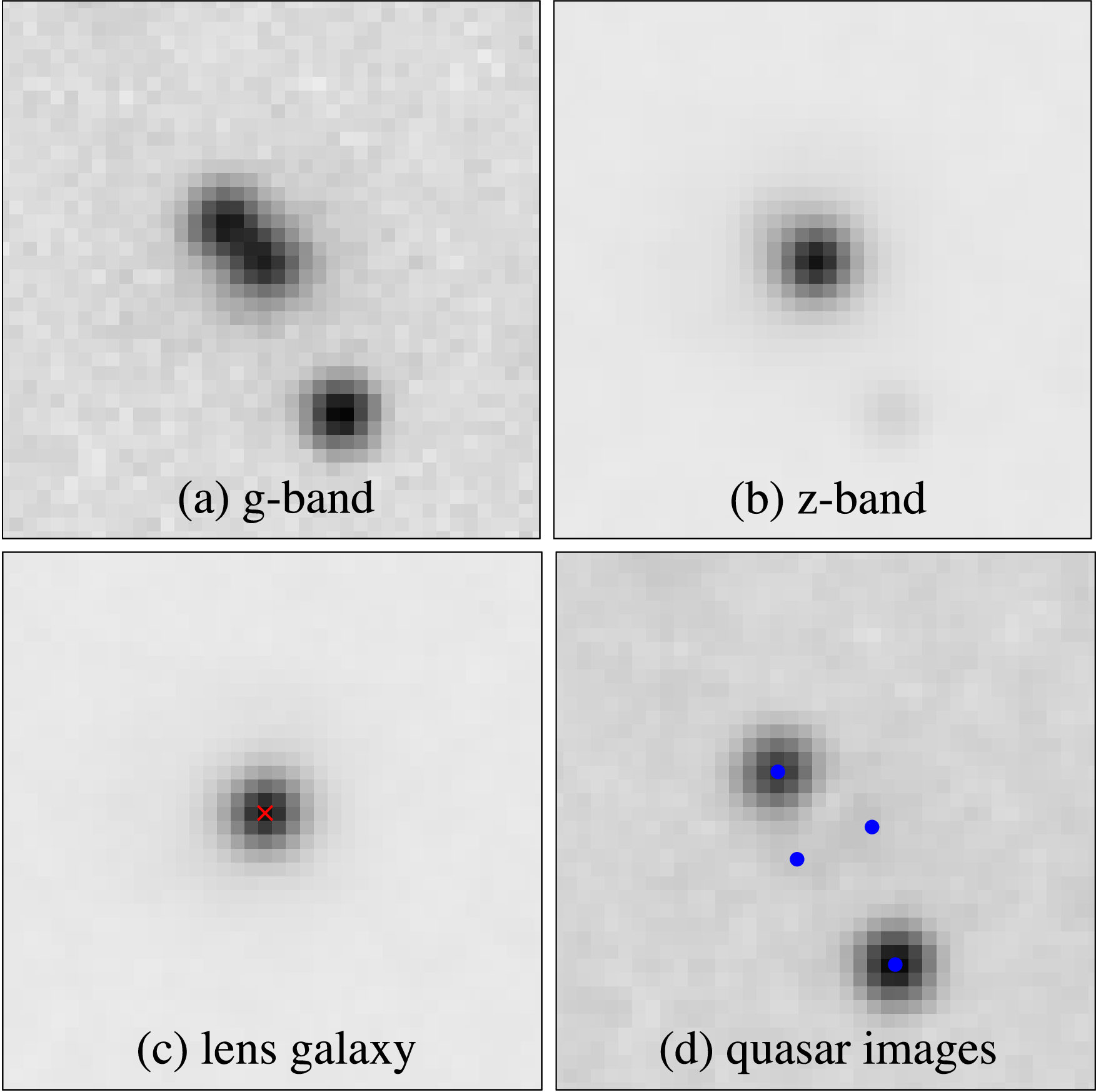}
\caption{An example of a simulated double system.  Panels (a) and (b):
  g-band and z-band cutouts, respectively.  
Panels (c) and (d): the lens galaxy and the quasar images,
respectively, which are separated based on color information and the
procedure described in \sref{subsec:separation}.  
The red cross in (c) is the estimated centroid of lens light.
When fitting four point sources to the quasar images in (d), the two quasar
images are correctly identified by the two blue dots, whereas the
remaining two blue dots are located at positions associated with noise peaks
or residuals due to imperfect lens-quasar separation.}
\label{fig:double_ex}
\end{figure}

There are four different scenarios of lensed objects depending on
their colors and brightnesses, and we
list the four cases in \tref{tab:case}. The most typical situation of
a lensed object is Case 1: quasar images are bluer and the lens galaxy
is brighter in the z-band. 
To build a versatile robot, we try to cover all four situations of lensed objects.
In the following subsections, we describe our procedure that works for
all cases, and illustrate
it with two typical examples of Case 1,  
one quad and one double, shown in \fsref{fig:quad_ex} and \ref{fig:double_ex}, respectively.

\subsection{Separation of lens and quasars}
\label{subsec:separation}

Since the color is different between the lens galaxy and quasar images,
below we describe a procedure to use cutouts of the lens
system in g-band and z-band to produce two images: one containing only
the lens galaxy, and another containing the lensed quasars only. 

a) Match up the point-spread function (PSF) in g and z bands, since the 
PSF is generally different between bands.
Specifically, we seek to determine the kernel function, $K$, such that
\begin{equation}
\label{equ:kernel}
\text{PSF}_g = K \ast \text{PSF}_z,
\end{equation}
where $\text{PSF}_g$ and $\text{PSF}_z$ are the PSFs in g-band and z-band, respectively. 
Here we assume the typical situation where $\text{PSF}_g$ is wider than $\text{PSF}_z$, but if the opposite is ture, we simply switch g and z in \eref{equ:kernel} and the following.
Based on the convolution theorem, the kernel function is then
\begin{equation}
\label{equ:kernelfft}
K = \text{IFT}\left\{{\text{FT}\{\text{PSF}_g\}\over \text{FT}\{\text{PSF}_z\}}\right\}, 
\end{equation}
where FT stands for the Fourier transform, and IFT is the Inverse Fourier transform.
However, in practice, numerical noise dominates in the high-frequency components 
since the ratio of $\text{FT}\{\text{PSF}_g\} / \text{FT}\{\text{PSF}_z\}$ is poorly behaved when $\text{FT}\{\text{PSF}_z\}$ is small. 
Therefore, we build a hybrid model for ${\rm FT}\{K\}$ where the low-frequency values are determined by $\text{FT}\{\text{PSF}_g\} / \text{FT}\{\text{PSF}_z\}$, and the high-frequency values are set to an elliptical Gaussian fit \citep{PhillipsEtal95} to the frequency components.
After obtaining the kernel function via the IFT, 
we can use \eref{equ:kernel} to match the PSFs in g-band and z-band; 
specifically, we convolve the band with the smaller FWHM with the kernel to match the
larger one of the other band.

b) Locate the brightest pixel in the z-band cutout, as illustrated in \fsref{fig:quad_ex}(b)
and \ref{fig:double_ex}(b), where we labelled them by $(i_{\rm max}, j_{\rm max})$. 
Typically, the lens galaxy is brighter than the quasar images in z-band (Case 1 and Case 3). 
Therefore, the brightest pixel in z-band is where the lens galaxy is located.
If Case 2 or Case 4 happens, the situation becomes the opposite, i.e.,
the brightest pixel in z-band is located at one of the quasar images.

c) Scale the brightest pixel value in z-band, $z(i_{\rm max}, j_{\rm max})$,
such that it becomes the
same as the value in g-band, $g(i_{\rm max}, j_{\rm max})$,
i.e., $g(i_{\rm max},j_{\rm max}) = \alpha z(i_{\rm max}, j_{\rm max})$, where $\alpha$
is the scaling factor. 
After subtracting pixel values in g-band from the scaled values in z-band, we obtain
\begin{equation}
\label{equ:qmap}
D_1(i,j) = g(i,j) - \alpha z(i,j),
\end{equation} 
where $i=1..N_{\rm x}$ and $j=1..N_{\rm y}$ are the pixel
indices in the image cutout of dimensions $N_{\rm x}\times N_{\rm y}$.
The image $D_1$ shows different outcomes for the four cases (\tref{tab:case}).
For Case 1, only quasar images are revealed in $D_1$, e.g., \fsref{fig:quad_ex}(d) and \ref{fig:double_ex}(d). 
When Case 2 happens, $-D_1$ represents the lens galaxy light distribution.
However, if quasar images are redder (Cases 3 and 4, which occur
less frequently), quasar images are revealed in $-D_1$ for Case 3 and the lens
galaxy light is defined by $D_1$ for Case 4.  We summarize the outcome
of image $D_1$ in \tref{tab:case}.

d) After obtaining either the quasar images or lens galaxies
successfully (i.e., image $D_1$) from the previous step, we can extract
the other component (i.e., the corresponding lens galaxies or quasar
images, respectively) with similar procedures as in b) and c). 
We identify the brightest pixel in $|D_1(i,j)|$ as $(i'_{\rm max},
j'_{\rm max})$ and scale the pixel value of $D_1(i'_{\rm max},j'_{\rm
  max})$ so that it is 
the same as that in the z-band $z(i'_{\rm max},j'_{\rm max})$,
i.e. $z(i'_{\rm max},j'_{\rm max})=\beta D_1(i'_{\rm max},j'_{\rm max})$, where
$\beta$ is the scaling factor. 
After calculating
\begin{equation}
\label{lmap}
D_2(i,j) = z(i,j)-\beta D_1(i,j),
\end{equation}
we show as examples the resulting image $D_2$ in \fsref{fig:quad_ex}(c) and
\ref{fig:double_ex}(c).
In \tref{tab:case} we summarize the outcome of $D_2$ for each case.

In this paper, we work with objects of Case 1 or 2, i.e., the quasar
images are bluer than the lens galaxy, which is the typical scenario
of lens systems.  To detect the rarer lens systems of Case 3 or 4,
one way is to first treat all objects as Case 1 or 2, classify them
(as described below in Sections \ref{subsec:identification} to
\ref{subsec:modelfit}), then treat all the failed Case-1 or Case-2 detections as
possible Case-3 or Case-4 candidates, and classify again.  This would allow
us to obtain candidates of all cases listed in \tref{tab:case}.

\subsection{Identifications of quasar image positions and lens center}
\label{subsec:identification}

After separating the quasar images and lens galaxy, we are able to obtain
$Q(i,j)$ and $L(i,j)$ from $D_1$ and $D_2$,
where $Q(i,j)$ is the image containing only the quasar's light and $L(i,j)$
contains only the lens galaxy's light (see \tref{tab:case}).
For probing an image configuration via modeling, we have to identify
image positions from $Q(i,j)$ and the lens center from $L(i,j)$. 

To identify the quasar image positions, we first adopt four point sources smeared by the matched PSF to obtain the
predicted image, $Q^{\rm P}(i,j)$. 
By varying the point source positions and brightnesses, we search for the
minimum difference between $Q^{\rm P}(i,j)$ and 
$Q(i,j)$ which is defined by 
\begin{equation}
\label{img}
\Delta Q^2 = \sum_{i,j} {[Q(i,j)-Q^{\rm P}(i,j)]^2}.
\end{equation}
Here we assume that the pixel uncertainty in $Q(i,j)$ is constant and
thus irrelevant in the minimization for the point source positions.
As shown in \fsref{fig:quad_ex}(d) and \ref{fig:double_ex}(d), we can
identify the image positions that are marked by the four blue dots. 
When there are only two images, two of the four dots would be located
at positions associated with remaining image residuals, or at random
positions when there are no significant residuals. 

To estimate the lens light centroid from the distribution $L(i,j)$, we 
calculate the first moments of $L(i,j)$.  The centroid (located at
fractional rather than integral pixels) is indicated by 
the red cross in \fsref{fig:quad_ex}(c) and \ref{fig:double_ex}(c).

\subsection{Potential quads and doubles via configuration}
\label{subsec:quadsdoubles}

\begin{figure}
\centering
\includegraphics[scale=0.4]{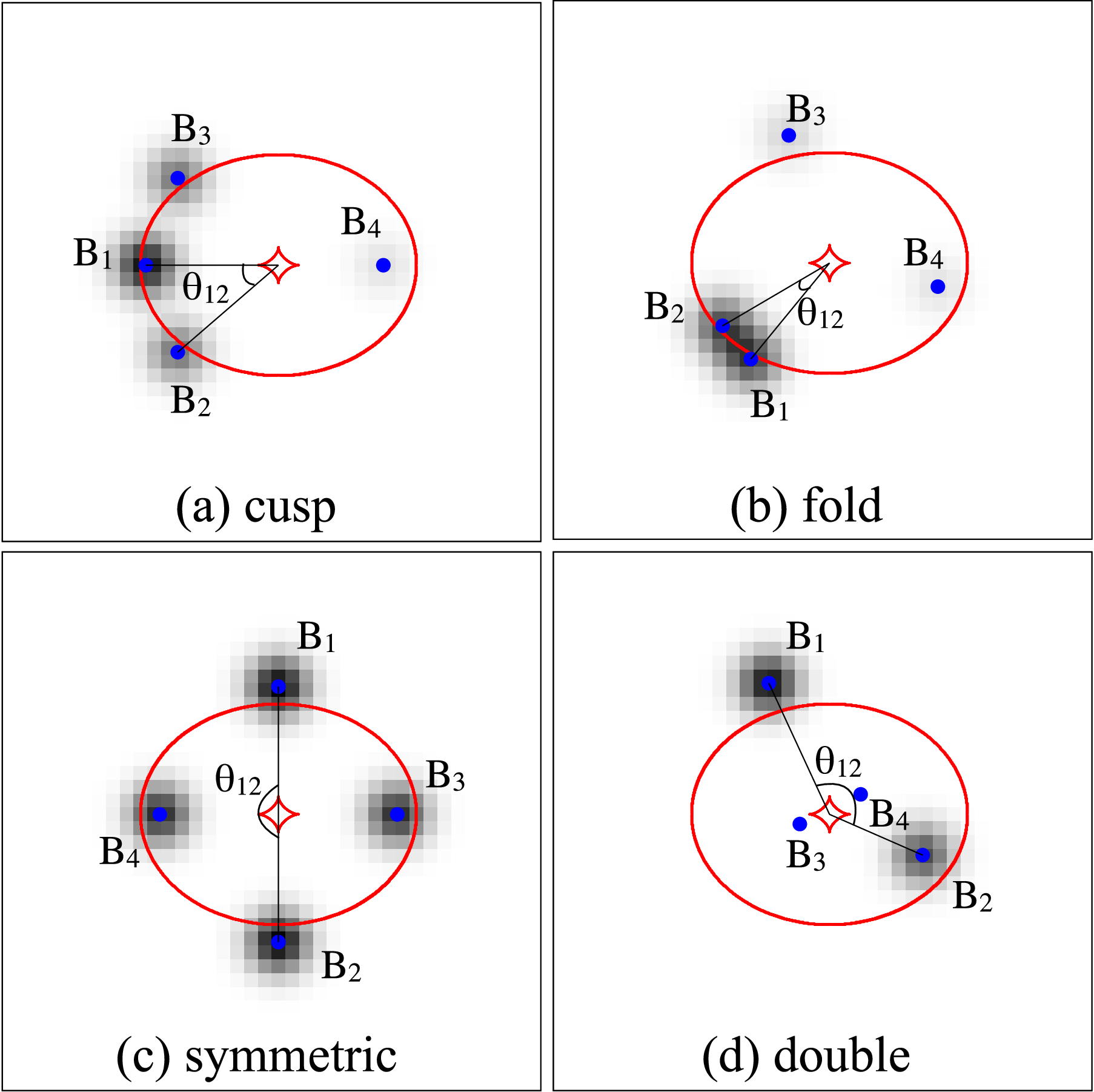}
\caption{Four generic configurations of strong lens systems: (a) cusp,
  (b) fold, (c) symmetric, and (d) double.  In each panel,
the brightnesses of four images are denoted by ${\rm B_1}$ to ${\rm
  B_4}$, in the order of decreasing brightness.
The angle formed by ${\rm B_1}$ and ${\rm B_2}$ with respect to the
lens center is labeled by $\theta_{\rm 12}$.
The red elliptical lines are critical lines where images are highly magnified. 
The red diamond-shaped lines are caustics where the critical
lines map to on the source plane.
Note that in (d) there are two additional images located at
noise-peak/residual positions.
} 
\label{fig:angle_config}
\end{figure}

We illustrate three generic image configurations of quads and one of
doubles in \fref{fig:angle_config}: (a) cusp, (b) fold, (c) symmetric, and (d) double. In
\sref{subsec:identification}, we mentioned that there are two dots
located at random/residual positions when fitting four dots to a double system.
When the quasar images can be well separated and there is no unrelated
object near the lens system,
the brightnesses of these two dots should be very faint. Therefore, we
can make use of this feature to 
classify potential quads and doubles.  Of the four identified dots, we denote $B_1$ as the brightest intensity value, $B_2$ as the second
brightest, $B_3$ as the third brightest, and $B_4$ as the
faintest. We further define $\theta_{12}$ as the angle subtended
between the locations of $B_1$
and $B_2$ with respect to the center of the lens galaxy (see \fref{fig:angle_config}).  Based on the generic image configurations of lenses, we classify objects with
$B_4/B_1 < 0.2$ and $\theta_{12} > 120^\circ$ as potential doubles,
and the remaining objects as potential quads.

\subsection{Classification via lens model fitting}
\label{subsec:modelfit}

\begin{figure}
\centering
\includegraphics[scale=0.3]{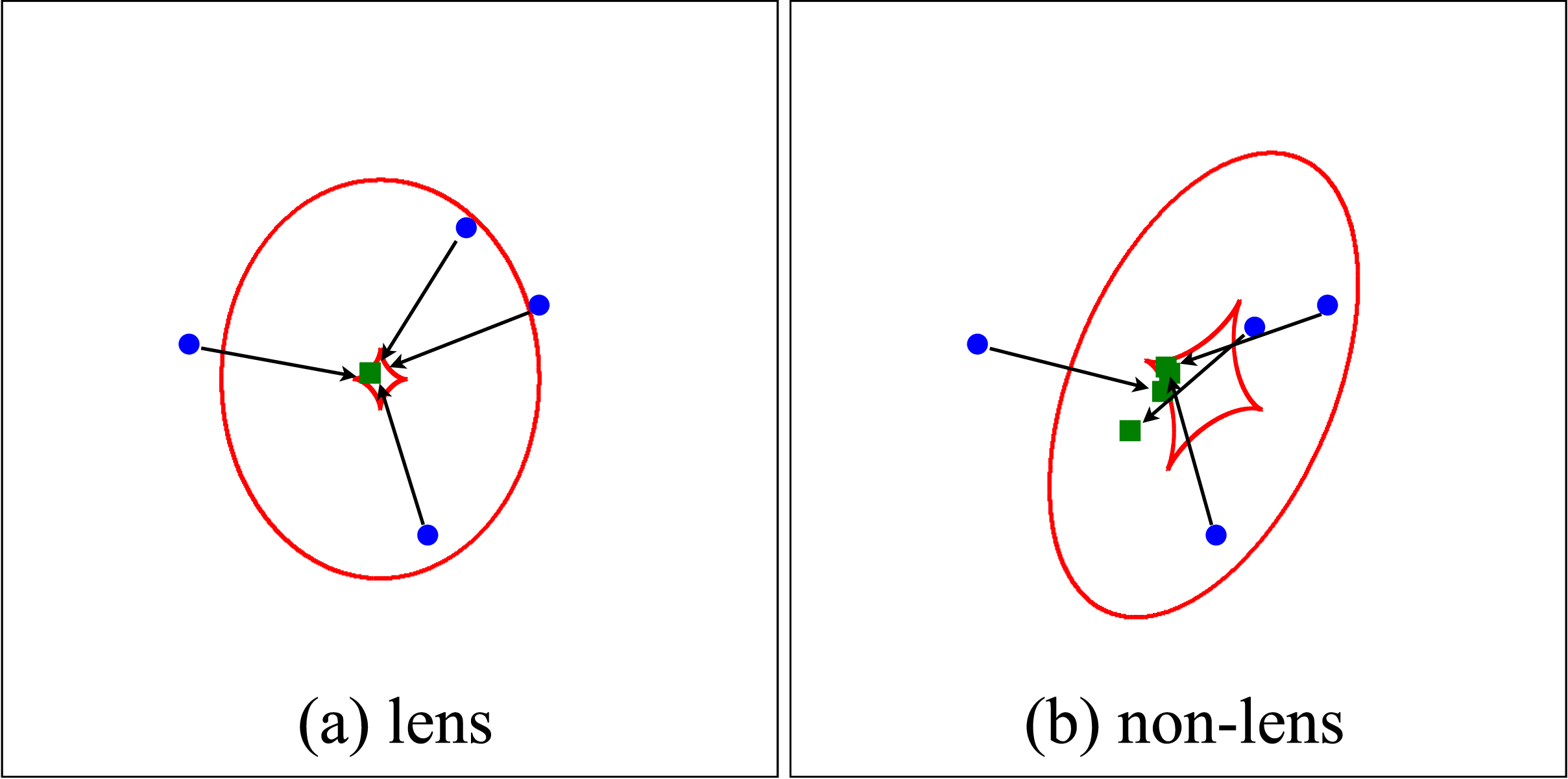}
\caption{Two examples to illustrate how \Chitah\ classifies lens
  (left) and non-lens (right) systems. In both panels, the red lines
  are the critical lines and caustics of the best-fitting lens model.
  The four blue dots indicate the quasar images, and the green squares
  are the mapped sources of the images from the best-fitting lens
  model.  We use the closeness of the mapped source positions to
  classify lens and non-lens systems: a system with quasar images that
  come from approximately the same source position is likely to be
  a lens (as in the left-hand panel (a)), whereas a system with quasar
  images that come from distinct source positions is likely to be a
  non-lens (as in the right-hand panel (b)).  }
\label{fig:lens_nonlens}
\end{figure}

After classifying the potential quads and doubles, we can model the
image configuations to detect plausible lens systems.  Specifically,
we try to see whether the quasar images could come from a single
source by varying a lens mass distribution centered close to the lens
light centroid.   
\fref{fig:lens_nonlens}(a) shows an example where we can construct a lens
mass model such that the quasar images could originate from the same
source.  In contrast, \fref{fig:lens_nonlens}(b) shows a configuration of
quasar images where we cannot find a lens model to make the quasar images
come from a single source.  Following \citet{MarshallEtal09}, we take
on the view that an object that can be well described by a lens model
is likely to be a lens.  Therefore, the lens model fitting illustrated
in \fref{fig:lens_nonlens} allows us to classify the left-hand object (a)
as a potential lens, and discriminate the right-hand object (b) as a
non-lens.  In the following, we describe in detail the lens model
fitting procedure. 

We model the lens mass distribution as an SIE profile, whose 2-dimensional surface mass
density is in the form 
\begin{equation}
\label{equ:kappa}
\kappa (x,y) = {\rein \over 2 \sqrt{x^2+y^2/q^2}},
\end{equation}
where $\rein$ is the Einstein radius, $q$ is the axis ratio, and $(x,
y)$ are the coordinates relative to the lens center.  Previous 
studies have shown that lens galaxies are close to having isothermal
profiles \citep[e.g.,][]{KoopmansEtal09, AugerEtal10, BarnabeEtal11,OguriEtal14}.  
The SIE is thus a simple profile that is adequate in describing
typical image configurations of lens systems.

We define the
 $\chi_{\text{src}}^2$ on the source (quasar) plane of the lens system as  
\begin{equation}
\label{equ:chi2src}
\chi_{\text{src}}^2= \sum_k{|{\bf r}_k-{\bf r}_{\text{model}}|^2\over\sigma_{\text{image}}^2/\mu_k},
\end{equation}
where ${\bf r}_k$ is the respective source position mapped from the
position of quasar image $k$ identified in $Q(i,j)$, $\mu_k$ is the
magnification at the position of quasar image $k$, and ${\bf r}_{\text{model}}$ is the
modeled source position evaluated as a weighted mean of ${\bf r}_k$,
\begin{equation}
\label{equ:src_model}
{\bf r}_{\text{model}} = {\sum_k \sqrt{\mu_k}{\bf r}_k \over \sum_k \sqrt{\mu_k}}
\end{equation}
\citep{Oguri10}. Here the index $k$ runs from 1 to 4 for the quad systems, and 1 to 2 for
the double systems. Since the quasar image positions are estimated through
minimizing \eref{img}, 
when adopting an imperfect PSF with a FWHM that varies by as much as
$0\farcs4$ (e.g., to account for possible PSF profile mismatch), the identified image positions could deviate by at most
$0\farcs2$.  Therefore, we adopt conservatively the uncertainty in the
identified quasar image positions,
$\sigma_{\text{image}}$, as $0\farcs2$. 

Since we can estimate the lens center from the light profile, it is 
useful to use it as a constraint on the center of the SIE lens mass
model because we expect the offset between the
light center and the mass center of isolated lenses to be small,
$\lesssim0\farcs05$ based on previous lensing studies 
\citep[e.g.,][]{KoopmansEtal06}.
Therefore, we define the $\chi_{\text{c}}^2$ as 
\begin{equation}
\label{equ:chi2c}
\chi_{\text{c}}^2= { |{\bf x}_{\text{model}}-{\bf x}_{\text{c}}|^2\over\sigma_{\text{c}}^2},
\end{equation}
where ${\bf x}_{\text{c}}$ is the lens center from the light profile, and
${\bf x}_{\text{model}}$ is the the lens center of the SIE model. 
Here we take $\sigma_{\text{c}}$ to be also $0\farcs2$ as an 
estimate of the uncertainty in identifying the lens
center from ground-based imaging.  

We define the total goodness of fit via
\be
\label{equ:chi2tot}
\chi^2 = \chi^2_{\text{src}}+\chi^2_{\text{c}}.  
\ee 
If we are able to find a lens model that can fit to the supposed
lensing features of an object (with a correspondingly small value of
$\chi^2$), then the object is likely to be a lens.  Thus, we can set a
threshold value, $\chith$, to decide between the lens and
non-lens classification: for $\chi^2 < \chith$, we classify
the object as a lens, and for $\chi^2 > \chith$, we classify
it as a non-lens.  In \sref{sec:results}, we explore the optimal value
for the threshold.

There are five parameters for the SIE model: $\rein$, $q$,
position angle, and lens coordinate ${\bf x}_{\text{model}}$. When
fitting to potential quad systems, there are four images to constrain
the model. However, when fitting to potential double systems, the two
images do not provide enough constraints on the SIE model, so we
choose the SIS model with three parameters (i.e., the spherical model
which eliminates $q$ and the position angle parameters).

\section{Simulation}
\label{sec:simulation}

\begin{figure*}[t]
\centering
\includegraphics[scale=1.]{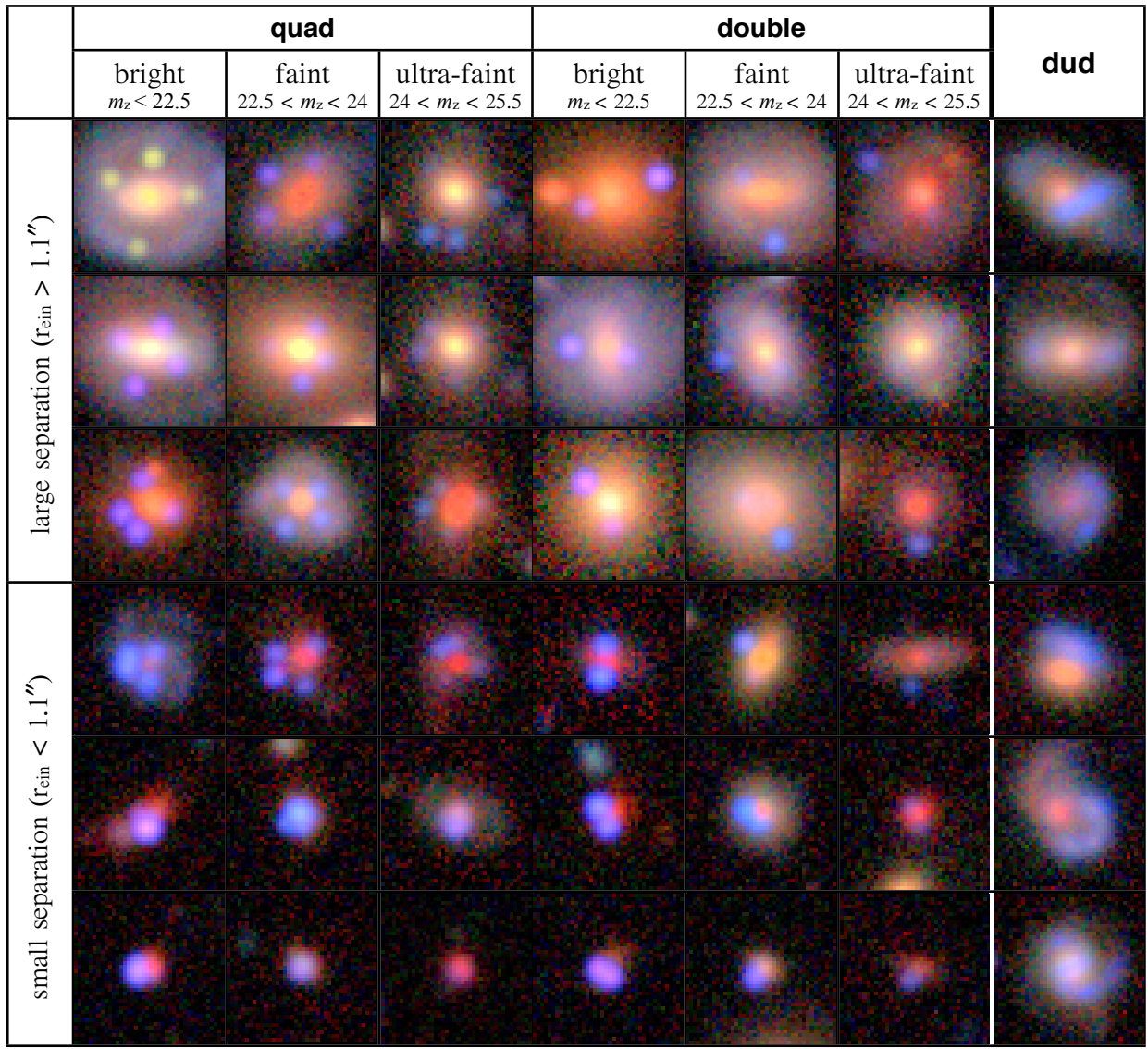}
\caption{Examples of the mock lenses (quads and doubles) with Gaussian PSF and duds used
  to test \Chitah.  Using the value of $\rein$ from the input
  SIE, we split the mock lenses into the ``large separation'' (with
  $\rein>1\farcs1$) and the ``small separation'' (with
  $\rein<1\farcs1$) samples.  Based on the z-band AB magnitude,
  $\zmag$, of the dimmest lensed quasar image, each of these samples is
  further divided into ``bright'' ($\zmag <22.5$), ``faint'' ($22.5
  < \zmag < 24$), and ``ultra-faint'' ($24 < \zmag < 25.5$).  In the
  last column we display examples of duds that are misidentified as
  possible lenses by citizen scientists in the Space Warps project.
  Each image cutout is $8''$$\times$$8''$.}
\label{fig:gallery}
\end{figure*}

To test the performance of \Chitah, we use the SIMCT\footnote{https://github.com/anupreeta27/SIMCT} code 
from Space Warps to generate a large sample of mock quasar lenses. Space 
Warps is a citizen science project that looks for lenses in imaging surveys via visual 
inspection \citep[][]{MarshallEtal15,MoreEtal15}.
Details of the SIMCT framework can be found in \cite{MoreEtal15}. 
Here, we briefly summarize the framework of SIMCT. 
The galaxy catalog from \cite{GavazziEtal14} is used to select massive and mostly early-type galaxies which are then parameterized as SIE lenses.
Using the SIE 
density profile with external shear (that accounts for external mass structures along 
the line of sight), point images of quadruply or doubly imaged lensed quasars are 
generated in the CFHTLS bands (ugriz) with realistic colors drawn from a quasar 
catalog. These point sources are subsequently blurred to match the image quality 
of the CFHTLS images (with PSFs of FWHM of $0\farcs8$ and $0\farcs7$ for the g-band and 
z-band, respectively). For the profile of the PSF, we consider two forms: (1) symmetric two-dimensional 
Gaussian with the aforementioned FWHM, and (2) Moffat described by
\begin{equation}
\label{equ:moffat}
\text{Moffat}(x,y)= {\beta - 1 \over \pi \alpha^2} \left[1+\left({x^2+y^2 \over \alpha}\right)^2\right]^{-\beta}, 
\end{equation}
where $\alpha$ and $\beta$ are seeing-dependent parameters, and ($x$, 
$y$) are the coordinates relative to the PSF center. 
We adopt $\alpha=0\farcs78$ and $0\farcs69$ for g-band and z-band, respectively, 
and $\beta=3.2$, which is the typical value obtained by fitting to the stars in CFHTLS (these $\alpha$ and $\beta$ values correspond to the aforementioned FWHM values).  After adding noise, these simulated quasar images are then 
superposed on top of the images of the real galaxies that were selected to be 
the lenses. In total, we use $\sim$$2000$ mock quads and $\sim$$3000$ mock doubles for each of the two forms of the PSF as the 
training set.  We note that our training set is different from the one described in 
\cite{MoreEtal15} in terms of the range of Einstein radii and the source 
magnitude limits. We explore a much wider range in these parameters here to test the 
performance of \Chitah.

Not only are simulated lenses needed, but false positives are also
important for coaching \Chitah. 
In this work, we employ 383 ``duds'' from Space Warps, which are
non-lensed objects that are misidentified as possible lenses by citizen
scientists.  These could be, for example, galaxies with several
point-like star formation regions around the bulge that could be
misidentified as quasar images, or the chance
alignment of point sources near a galaxy.

We show some examples of the mock lenses with the Gaussian PSF and the duds in
\fref{fig:gallery}.  We categorize the mock lenses into six groups
based on the quasar image separation and brightness. 
When the input $\rein$ of the mock lens is larger or smaller than
1\farcs1, we classify the lens system as large- or small-separation
lens, respectively.
Furthermore, we use the magnitude in the z-band of the dimmest image,
$\zmag$, to separate the mock lenses into three categories:
``bright,'' ``faint,'' and ``ultra-faint,'', corresponding to $\zmag
<22.5$, $22.5 < \zmag < 24$, and $24 < \zmag < 25.5$, respectively.


\section{The performance of \Chitah}
\label{sec:results}

We investigate the performance of \Chitah\ in classifying the duds and
simulated lenses described in the previous section.  We first consider
the simulations produced with the Gaussian PSF in
\sref{sec:results:gaussian}, and explore the effect of the PSF in
\sref{sec:results:moffat} by analyzing the simulations with Moffat
PSFs.

\subsection{Simulations with Gaussian PSF}
\label{sec:results:gaussian}

\begin{figure}
\centering
\includegraphics[scale=0.2]{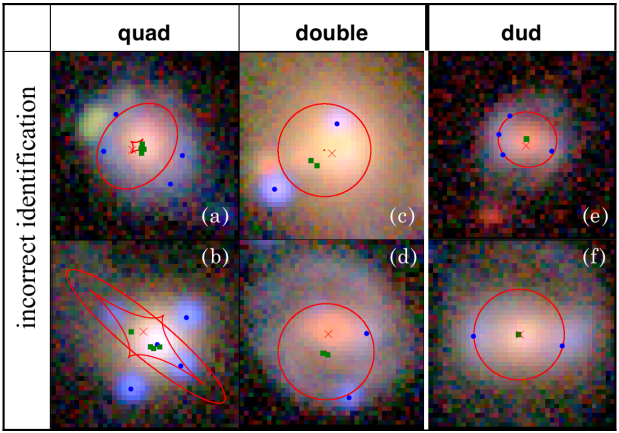}
\caption{Examples of the incorrect identifications.  Mock quads, mock
  doubles and duds are shown in the left, middle and right columns
  respectively. The mocks are generated with Gaussian PSFs. Each image cutout is
$8''$$\times$$8''$.  The red cross is the estimated centroid of the lens light. We 
identify the locations of quasar images, which we indicate with the 4 blue dots. 
The red elliptical lines are critical lines where lensed images are highly magnified. 
The red diamond-shaped lines are caustics where the critical lines map to on the source plane. 
The green squares are the mapped sources of the identified quasar
images from the best-fitting lens model. 
Panel (a) shows that \Chitah\ mis-identifies the four image positions
as the blue lens light residuals rather than the faint quasar images
that are in green.
Panel (b) shows that \Chitah\ mis-identifies one quasar image because of imperfect lens-quasar separation. 
Panel (c) shows that a large $\chi^2$ results from the two quasar
images not being collinear with the lens light center, leading to an
incorrect classification of this system as a non-lens.
Panel (d) shows that \Chitah\ mis-identifies the position of the fainter quasar image.
Panels (e)/(f) show that the surrounding blue blobs are misidentified
as quasar images that can be well fitted by an SIE/SIS model by chance.
}
\label{fig:falseclassification}
\end{figure}

\begin{figure*}[t]
\centering
\includegraphics[scale=0.65]{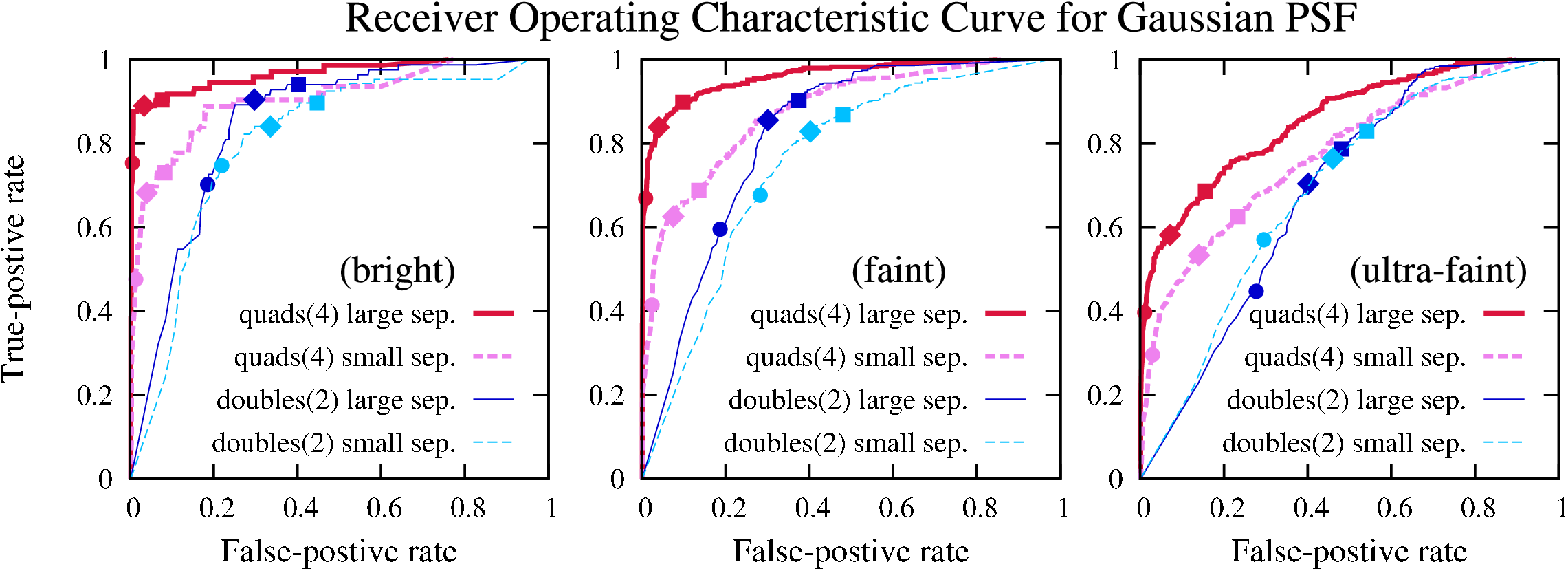}
\caption{ROC curves for the mock lenses with Gaussian PSF. 
The three brightness groups are formed based on $\zmag$ of the dimmest quasar image:
bright (left-hand panel) with $\zmag < 22.5$, faint (middle panel)
with $22.5 < \zmag < 24$, and ultra-faint (right-hand panel) with
$24 < \zmag < 25.5$. 
The thick solid and thick dashed curves display the results of mock quad
lenses with large ($\rein$ $>$ 1\farcs1) and small ($\rein$ $<$
1\farcs1) quasar image separations, respectively.
Similarly, the thin solid and thin dashed curves show the results of
mock double lenses with large and small separations, respectively.
Each curve is obtained by plotting TPR vs. FPR for various
$\chith$ settings.
The locations of $\chith = 1, 4$ and $7$ are indicated by
circles, diamonds, and squares, respectively, on each curve.
\Chitah\ is able to capture bright quads with large separations 
with TPR $\sim$ 90\% and FPR $\sim$ 3\% when $\chith \sim 4$,
and bright doubles with large separations with TPR $>$ 70\% and
FPR $<$ 20\% when $\chith \sim 1$.  Large-separation quads
are easier to detect than small-separation quads. 
}
\label{fig:roc_curve_gau}
\end{figure*}

\begin{figure*}[t]
\centering
\includegraphics[scale=0.65]{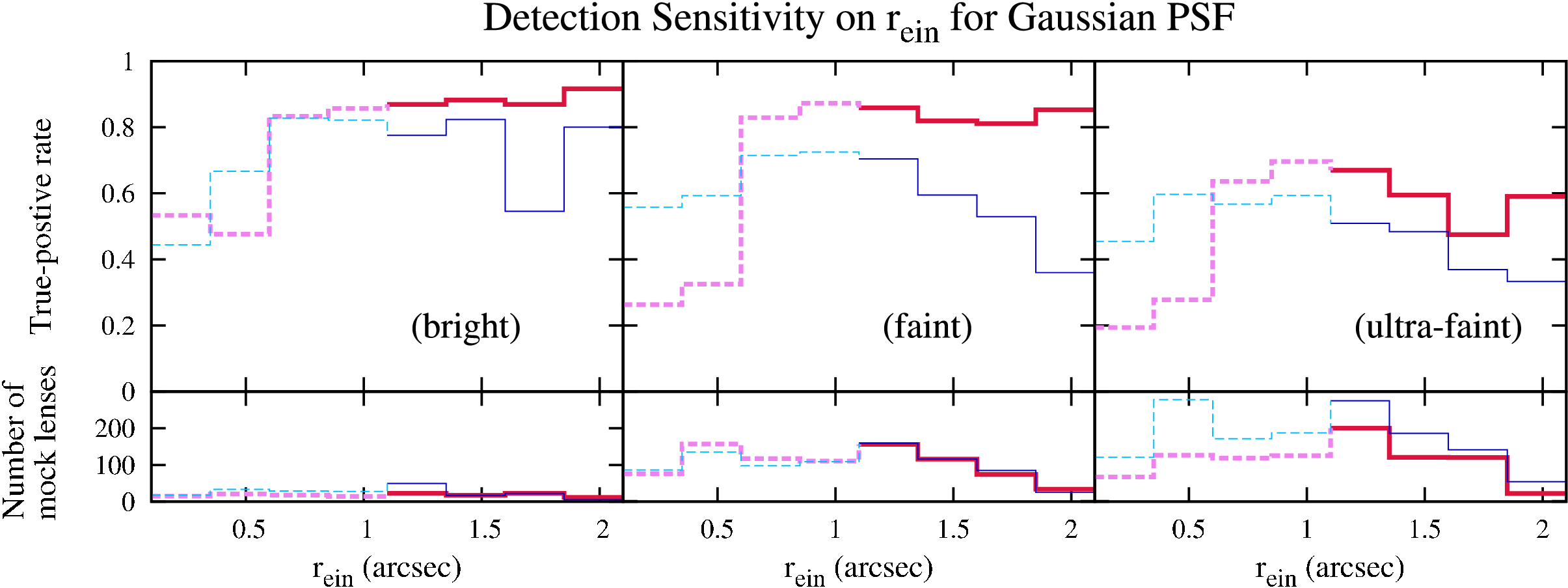}
\caption{Dependence of mock-lens detection with Gaussian PSF on the input $\rein$.
The three brightness groups are formed based on $\zmag$ of the dimmest quasar image:
bright (left-hand panel) with $\zmag < 22.5$, faint (middle panel)
with $22.5 < \zmag < 24$, and ultra-faint (right-hand panel) with $24 < \zmag < 25.5$. 
The top panels show the detection sensitivity on $\rein$,
and the bottom panels are the corresponding number of mock
lenses for each $\rein$ bin. 
The TPR for each bin is estimated at $\chith = 4$ for
quads and $\chith = 1$ for doubles.  
\Chitah\ can identify quads (thick lines) robustly for $\rein\gtrsim0\farcs5$,
corresponding to the limit set by the image quality of the simulated
mocks with the PSF FWHM of $0\farcs8$. There is a gradual decline in the TPR
for the doubles (thin lines) as $\rein$ increases due to a tendency for the lens
mass center to be not collinear with the two quasar images for large
$\rein$ (arising 
from either lens ellipticity or external shear), resulting in the SIS
model failing to identify such objects as lenses.  
}
\label{fig:detection_rein_gau}
\end{figure*}

To quantify the performance of \Chitah, we plot the receiver operating characteristic (ROC) curve: the relation between true- and false-positive rates.
The definitions of true-positive rate (TPR) and false-positive rate (FPR) are
\begin{equation}
\label{equ:tpr}
\text{TPR} = {\text{\small \# of correct identifications of positive instances} \over
  \text{\small \# of positive instances}},
\end{equation}
and
\begin{equation}
\label{equ:fpr}
\text{FPR} = {\text{\small \# of incorrect identifications of negative instances} \over
  \text{\small \# of negative instances}}.
\end{equation}
We can quantify individual TPR and FPR for the quads and doubles given
the mock quads, mock doubles and, duds that we have from
\sref{sec:simulation}.  For the quads, the number of positive
instances is the number of mock quads, whereas the number of negative
instances is the number of non-quads, which is the sum of the numbers
of duds and mock doubles.  Similarly, for the doubles, the number of
positive instances is the number of mock doubles, and the number of
negative instances is the sum of the number of duds and mock quads.

We have previously illustrated correct identifications in
  \fref{fig:lens_nonlens}.  Here, we show examples of incorrect
  identifications in 
\fref{fig:falseclassification} for $\chi^2_{\rm
  th}$ (see \sref{subsec:modelfit}) of 4 for quads and 1 for doubles.
In each panel, the lens light centroid and the multiple quasar images that are identified by \Chitah\ are indicated with
a red cross and blue dots, respectively. The green squares are the
mapped source positions of the quasar images from the best-fitting lens
model whose critical lines and caustics are shown as red elliptical lines and diamond-shaped lines, respectively.
We describe the reason for the incorrect identification in each panel
of \fref{fig:falseclassification} as follows:
(a) the four quasar images are misidentified at the blue lens
    light residuals rather than at the locations of the green blobs
    that are the simulated quasar images;
(b) one quasar image is misidentified at the lens residual
    near the lens center due to imperfect lens-quasar separation;
(c) the large $\chi^2$ (hence the incorrect non-lens classification) results from the two quasar images not being collinear with the lens light center;
(d) one faint quasar image (near the top of the lens galaxy)
    is misidentified at a blue starforming region within the spiral
    arms of the lens galaxy;
(e) the surrounding blue ring of a galaxy is misidentified as
    quasar images that can be well fitted by an SIE model by chance
    (i.e., incorrectly identified as a quad lens); 
(f) the two blue star-forming regions are misidentified as
    quasar images that are well fitted by an SIS model by 
    chance (i.e., incorrectly identified as a double).

For a given value of the threshold $\chith$ to classify
between lens and non-lens (see \sref{subsec:modelfit}), 
we can compute the TPR and FPR of the mock
lenses and duds.  A larger $\chith$ threshold leads to
both higher TPR and FPR.  The reason is that it is easier to find a
lens model to map the quasar positions with a $\chi^2$ value (in
\eref{equ:chi2tot}) less than the threshold $\chith$ when
$\chith$ is large, and hence the higher TPR.  At the same
time, a large $\chith$ also means that we can fit a
non-lens more easily with a lens model, resulting in a higher FPR.

As shown in \fref{fig:roc_curve_gau}, we plot the
ROC curves for each of the ``bright'' (left-hand panel), ``faint''
(middle panel), and ``ultra-faint'' (right-hand panel)
samples.  Each curve is mapped out by varying $\chi^2_{\rm th}$: we
start at the lower-left corner of the plot with a small $\chi^2_{\rm
  th}$ value, and as we increase $\chi^2_{\rm th}$, we go along the
curve toward the top-right corner.  The goal is to be near the
top-left corner with a high TPR and a low FPR.  In each curve, we mark
the locations of $\chith = 1, 4$ and $7$ by circles, diamonds, and
squares, respectively. For the bright quads with large separations
(thick solid curve in the left-hand panel), \Chitah\ is able to
capture these quads with a TPR $\sim$ 90\% and FPR $\sim$ 3\% when
$\chith \sim 4$. Even for the faint quads with large separations (thick solid curve
in the middle panel), we obtain TPR  $>$ 80\% and FPR $<$ 5\% when $\chith \sim 4$. 
In general, large-separation quads (thick solid curves) are easier to identify than
small-separation quads (thick dashed curves) given the higher ROC curves
of large-separation quads.
Also, the ROC curves of quads are closer to the top-left corner than those of doubles.
This implies that \Chitah\ can hunt down a much purer sample of quad candidates
than double candidates. According to the ROC curves, we can adopt the appropriate threshold for 
quad and double classifications, \ie\ $\chith\sim4$ for quads and
$\chith\sim1$ for doubles.  We expect such threshold values to be
applicable to imaging surveys that have image qualities similar to
that of our mock lenses based on CFHTLS.
For surveys whose image quality is different from CFHTLS, 
one can first simulate realistic mocks for these surveys, and then make the ROC curve to choose an appropriate $\chith$.  
We note that the $\chith$ estimated from such simulations serves as a good guid and can be tuned 
when applying to real data -- for example, 
one could first choose a low $\chith$ to get the most probable candidates 
when searching through an imaging survey, and then gradually relax/increase $\chith$ to get more candidates that are likely less pure.  
Moreover, one can also test \Chitah\ on a smaller region of the actual survey, which covers previously well studied fields and
where there are known lensed quasars. This will give us an idea of which $\chith$ value provides an optimal balance between completeness and purity,
and we can then apply such $\chith$ for the entire survey.
  
In \fref{fig:detection_rein_gau}, we investigate the detection sensitivity on
$\rein$ (which is roughly half of
the quasar image separation).  The top panels show the TPR that is
estimated with $\chith = 4$ for quads and $\chith = 1$ for doubles.  The number of mock
lenses for each $\rein$ bin is shown in the bottom panels.  
As seen from the top panels, \Chitah\ is able to capture
quad lenses with large $\rein$ with nearly constant TPR as set by
$\chith$.  However, we see a sharp drop in TPR as $\rein$
becomes smaller than $0\farcs5$.  Small-separation lenses ($\lesssim1''$ with
$\rein<0\farcs5$) are harder to detect since the quasar images are blended
together given the PSF FWHM of $0\farcs8$.  Therefore, the performance
of \Chitah\ in detecting small-separation quads is set by the
image quality.
For the doubles (thin lines), the TPR shows a decline at both small
$\rein$ and large $\rein$.  At small $\rein$, it is more difficult to
resolve the two quasar images, so it is harder to fit an SIS model.
However, the drop 
in TPR for doubles is not as drastic as in that of the quads because,
with only two images, it is relatively easy to use an SIS to constrain the
image configurations.  Note that mock doubles with hifh TPR also
have correspondingly high FPR ($>20\%$), as is visible in \fref{fig:roc_curve_gau}. At $\rein\gtrsim 1''$, there is
also a gradual decline in the TPR for the doubles as $\rein$
increases.  This is due to the typically larger offset between the SIS
centroid and the light centroid of the lens galaxy as the quasar image separation
increases.  The input mass distribution for generating the quasar image
configuration is an SIE with external shear, which could lead to the
two quasar images not being collinear with the lens mass center.  The
offset is typically larger for doubles with larger quasar 
image separations (i.e., larger $\rein$).  In contrast, the SIS model by
construction has its mass center collinear with the two 
predicted quasar image positions.  Therefore, the SIS model will tend to
produce higher $\chi^2_{\rm c}$ in \eref{equ:chi2tot} for larger $\rein$, causing a
decline in the TPR.  For the quads, the decline in TPR at large
$\rein$ is less apparent because we use an SIE model to fit to the
quad configuration and the effect of the external shear can be mostly
absorbed into a change in the ellipticity of the SIE to yield a low
$\chi^2$.

\subsection{Simulations with Moffat PSF}
\label{sec:results:moffat}

\begin{figure*}[t]
\centering
\includegraphics[scale=0.65]{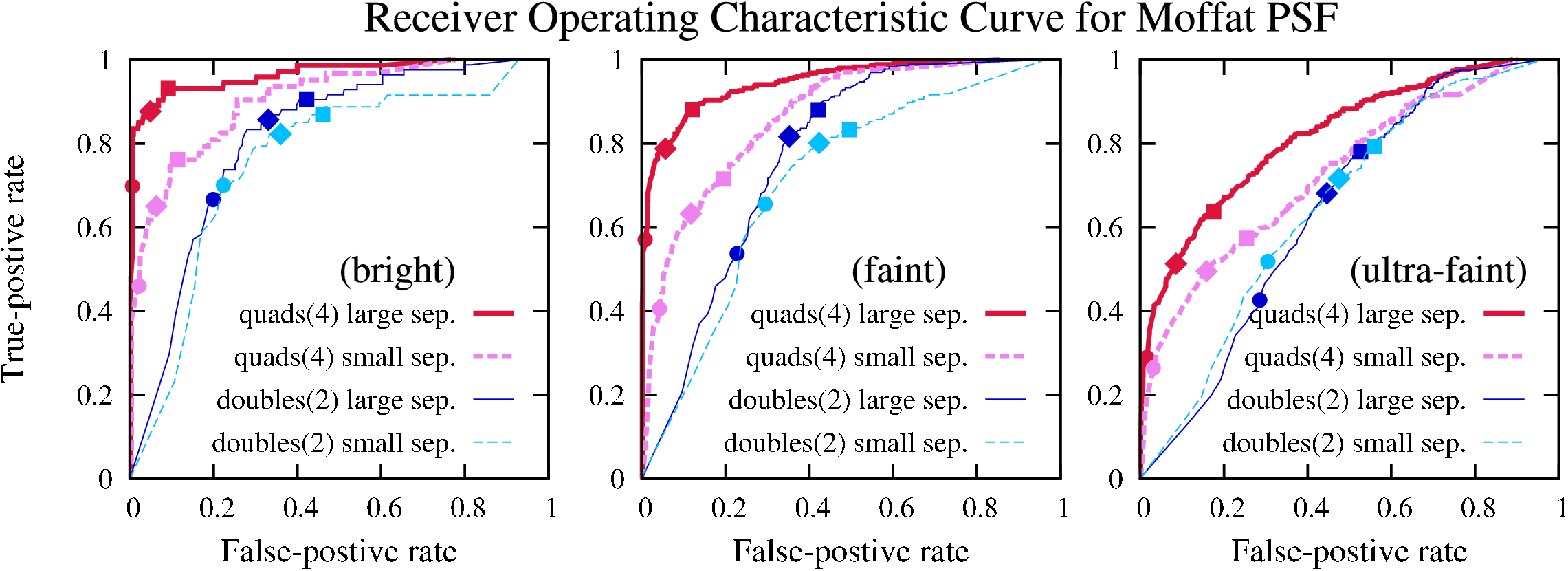}
\caption{ROC curves for the mock lenses with Moffat PSF. 
The panels are similar to those for the Gaussian PSF mocks (\fref{fig:roc_curve_gau}).
\Chitah\ is able to capture bright quads with large separations 
with TPR $\sim88\%$ 
 and FPR $\sim5\%$ 
when $\chith \sim 4$,
and bright doubles with large separations with TPR $>$ 66\% and
FPR $<$ 20\% when $\chith \sim 1$.  
In comparison with the ROC curves for Gaussian mocks, the Moffat PSF leads to a lower ROC curve at the level of a few percent.
}
\label{fig:roc_curve_mof}
\end{figure*}

\begin{figure*}[t]
\centering
\includegraphics[scale=0.65]{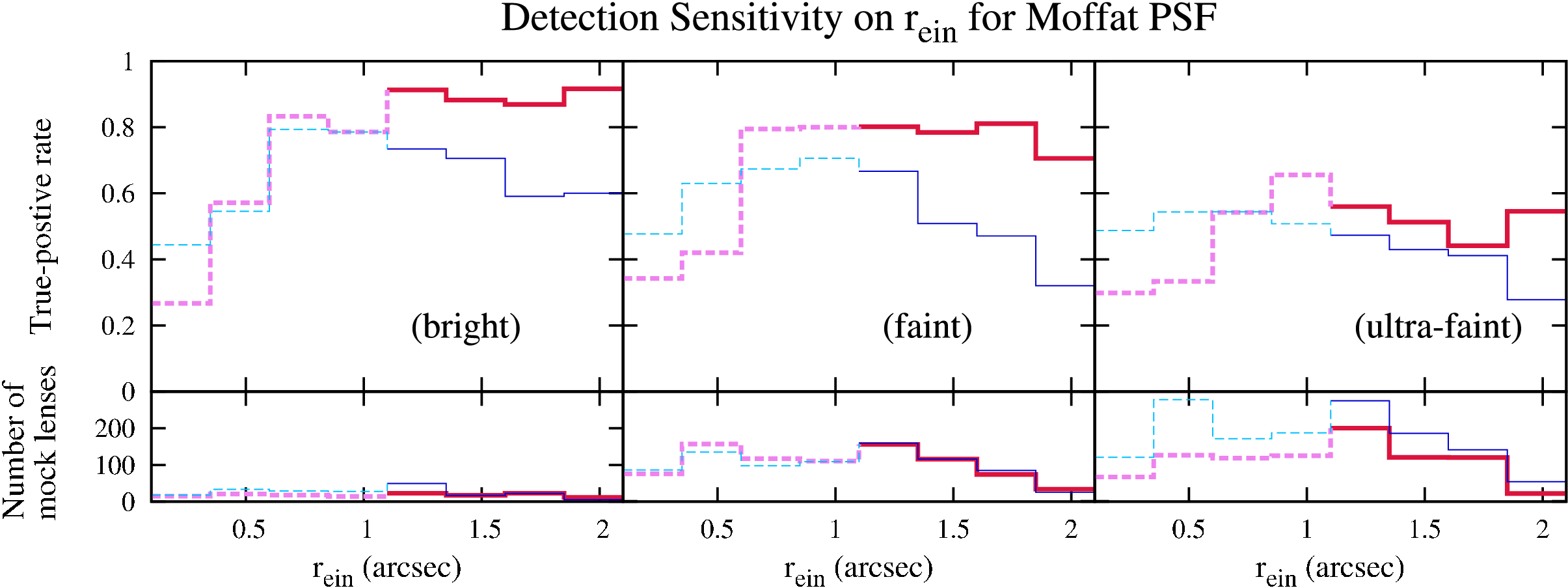}
\caption{Dependence of mock-lens detection on the input $\rein$ for mocks with Moffat PSF.
The three brightness groups are formed based on $\zmag$ of the dimmest quasar image:
bright (left-hand panel) with $\zmag < 22.5$, faint (middle panel)
with $22.5 < \zmag < 24$, and ultra-faint (right-hand panel) with $24 < \zmag < 25.5$. 
The top panels show the detection sensitivity on $\rein$,
and the bottom panels are the corresponding numbers of mock
lenses for each $\rein$ bin. 
The TPR for each bin is estimated at $\chith = 4$ for
quads and $\chith = 1$ for doubles.  
The panels are similar to those for Gaussian PSF mocks, but TPRs decrease by $\sim5\%$ due to the more extended wings of the Moffat PSF.
}
\label{fig:detection_rein_mof}
\end{figure*}

To further explore the effect of the PSF, we generalize the Gaussian PSF to the Moffat function defined in \eref{equ:moffat}, 
which describes better the wings of PSFs in CFHTLS observations. 
After running \Chitah\ on the same data set of mocks with Moffat PSF, we plot the ROC curves and the detection sensitivities of $\rein$ 
for each sample as shown in \fref{fig:roc_curve_mof} and \fref{fig:detection_rein_mof}, respectively.

The results for Moffat PSF have very similar features to those for Gaussian PSF, although slightly worse due to the more extended wings of the Moffat PSF. 
For the mocks with Moffat PSF, \Chitah\ is able to reach TPR $\sim88\%$ and FPR $\sim5\%$ when $\chith\sim4$ 
for the bright quads with large separations (thick solid curve in the left-hand panel).
As a result of the extended PSF wing effect, we notice that the ROC curves for Moffat PSF become slightly lower. 

The TPRs decline by $\sim2\%$ for the bright quasar sample, and by $\sim7\%$ for the ultra-faint quasar sample.
Based on this simple test, it appears that the form of the PSF has an effect on the ROC at the level of a few percent.  
We also note the importance of having good PSF models in imaging suveys in order to match the PSFs of different bands.  
In short, \Chitah\ performs best for quad lens systems with bright and widely separated quasars and with narrow and well characterized PSFs.

\section{Application to \anguita}
\label{sec:application}

\begin{figure}
\centering
\includegraphics[scale=0.4]{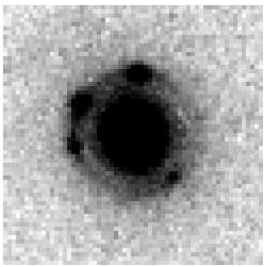}
\caption{The \hst ACS F814W exposure cutout
  image of \anguita, discovered by \citet{FaureEtal08}. The elliptical galaxy at the center is the lens galaxy, and 
four lensed quasar images are located around the lens. The pixel scale
is 0\farcs05 and the field of view is $3''$$\times$$3''$. 
The cutout image is obtained from the data release of \citet{FaureEtal08}. }
\label{fig:cosmos5921}
\end{figure}

To demonstrate that \Chitah\ not only captures simulated lenses
but also real lenses, we consider the known gravitational lens
\anguita\ in this section. 

\subsection{Observations of \anguita}

The lens system \anguita\ is one of the 67 strong lens candidates
discovered by \citet{FaureEtal08} via visual inspection of early-type
galaxies with redshifts $< 1.0$ in the $1.64 \deg^2$ \hst\ COSMOS
survey \citep{ScovilleEtal07}.  
The \hst\ Advanced Camera for Surveys (ACS) F814W exposure of \anguita,
obtained from the data released by
\citet{FaureEtal08}\footnote{http://wwwstaff.ari.uni-heidelberg.de/mitarbeiter/cfaure/cosmos\\/info/info.5921+0638.html}, 
is shown in \fref{fig:cosmos5921}. \citet{AnguitaEtal09} obtained spectroscopic follow-up observations and
performed a detailed analysis of the lensing system \anguita.  They
have confirmed that \anguita\ is a lensed quasar and not a lensed galaxy
based on the morphology, i.e, the four point-like images that lie
around an early-type galaxy suggest that the background source is a
quasar.  They also measured the lens redshift to be $\zl = 0.551$ from
FORS1 observations and inferred a possible AGN source redshift of
$\zs=3.14$ from the u* drop-out criterion and a candidate Ly-alpha line.
\citet{AnguitaEtal09} found that an SIE with a small amount of external shear
($\gamma = 0.038$) provides an adequate fit to the observed
positions of the quasar images in \anguita.  

We attempt to feed \Chitah\ this lensed quasar system to test its
ability.  We use the ground-based images of \anguita\ from the
Suprime-Cam on the 8.2m Subaru Telescope \citep{TaniguchiEtal07}.  The
optical images are obtained in six broadbands: $B$, $g'$, $V$, $r'$,
$i'$ and $z'$. We obtained the images from the Subaru archive and
reduced the images using the HSC pipeline, a derivative of the LSST
pipeline\footnote{https://confluence.lsstcorp.org/display/LSWUG/LSST+Software\\+User+Guide}\citep{IvezicEtal08, AxelrodEtal10},
modified for use with Suprime-Cam and Hyper Suprime-Cam.  
The images were overscan-subtracted,
flat-fielded using the COSMOS
flats\footnote{http://irsa.ipac.caltech.edu/data/COSMOS/images/subaru/flats}
and calibrated against the Eighth Data Release of the Sloan Digital
Sky Survey \citep{AiharaEtal11}.  These calibrated images were warped
to a common coordinate system and combined with mild clipping ($3\sigma$, 2
iterations) clipping to reject extreme outliers.  The aperture
photometry on the coadd has an RMS difference of $\sim$4\% in r-band
against SDSS, and the astrometry RMS difference is 100 mas.

The $B$- and $z'$-bands image cutouts are illustrated in
\fsref{fig:cosmos5921_ex}(a) and (b), respectively.  The pixel scale is
$0\farcs20$.
We estimate the PSF FWHM to be $\sim 0\farcs55$ for $B$-band and $\sim
0\farcs8$ for $z'$-band by fitting symmetric two-dimensional Gaussians to stars in the field.

\subsection{\Chitah\ on \anguita}

\begin{figure}
\centering
\includegraphics[scale=0.4]{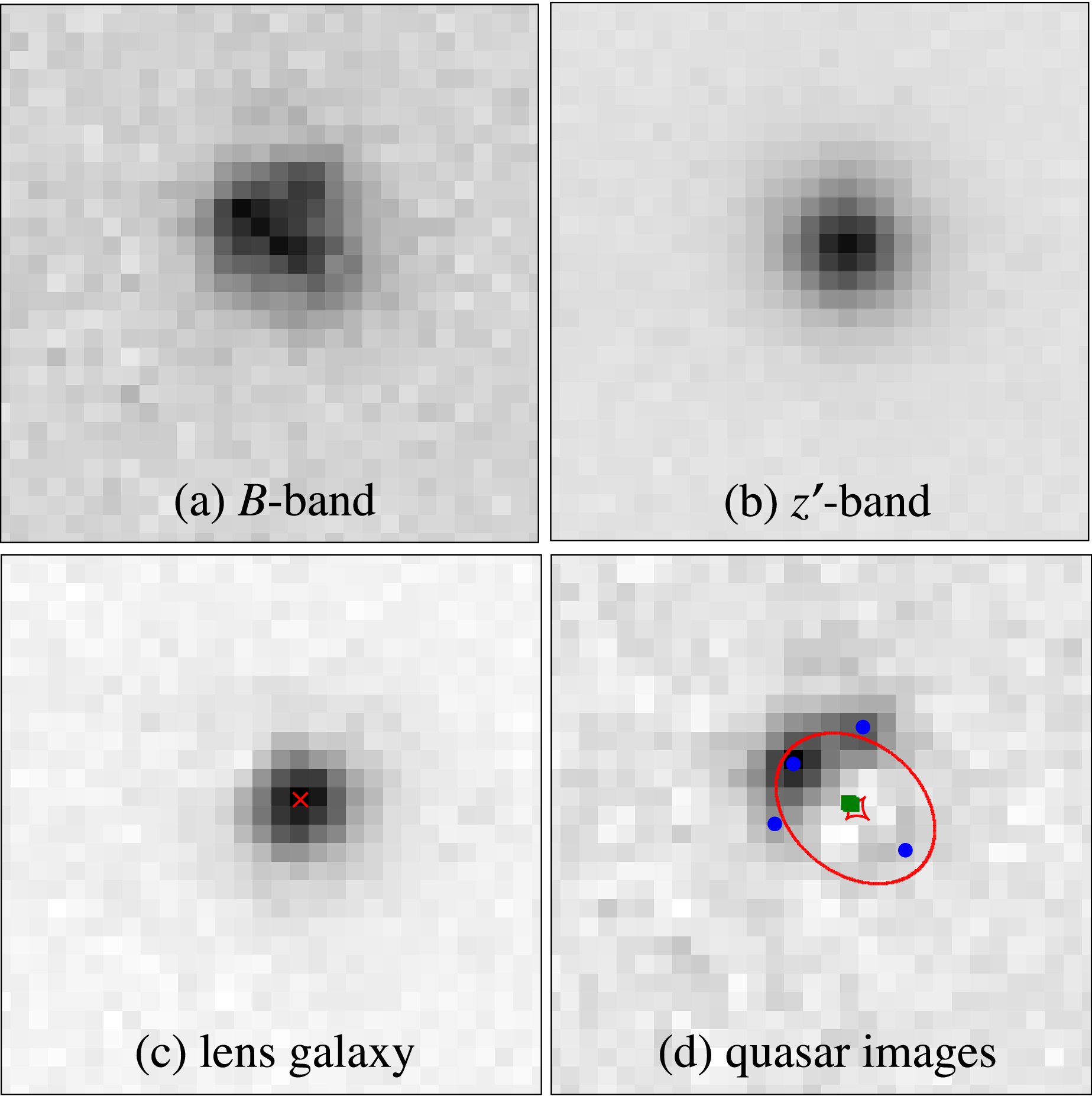}
\caption{Classification of \anguita\ by \Chitah. Panels (a) and (b): $B$- and $z'$-band cutouts, respectively.
Panels (c) and (d): the lens galaxy and the quasar images,
respectively, which are separated based on color information and the
procedure described in \sref{subsec:separation}.
The red cross in (c) is the estimated centroid of lens light.
We identify the locations of quasar images, which we indicate with the
4 blue dots in (d). The red elliptical lines are \textit{critical lines} where images are highly magnified.
The tiny red diamond-shaped lines near the center are \textit{caustics} where the critical
lines map to on the source plane. The green squares are the mapped sources of each image from the best-fitting lens model.
The closeness of all the mapped sources indicates that \anguita\ is
indeed a lensed system.}
\label{fig:cosmos5921_ex}
\end{figure}

To separate the lens and images more effectively, we choose the cutouts 
in the bluest band and in the reddest band, i.e. the $B$- and $z'$-bands.
The performance of \Chitah\ is illustrated in
\fref{fig:cosmos5921_ex}.  We can separate cleanly the
lens galaxy light (panel (c)) and quasar images (panel (d)).
We can also classify \anguita\ as Case 1 (where the quasar images are
bluer than the lens galaxy and the lens galaxy is brighter than the
quasar images in the $z'$-band).
After separation of lens and quasars, we are able to identify image
positions labeled by the four blue dots in \fref{fig:cosmos5921_ex}(d). 
The lens centroid is also estimated and labeled as a red cross in \fref{fig:cosmos5921_ex}(c). 
We can fit an SIE model to the observed image configuration; 
as shown in \fref{fig:cosmos5921_ex}(d), the mapped sources from the
quasar images lie close together with a $\chi^2 \sim
0.53$, and the Einstein radius of the best-fit SIE model is $\rein \sim0\farcs74$, 
which is comparable to the Einstein radius ($0\farcs71$) measured by
\cite{AnguitaEtal09}. 
To categorize \anguita, we estimate the signal-to-noise ratio (SNR)
of the faintest quasar image in $z'$-band/z-band for both COSMOS and CFHTLS
based on the smeared $\zmag$ brightness, 
because the COSMOS depth differs from that of the CFHTLS.  
The SNR of \anguita\ is $\sim$$8$, which can be classified in the
faint group, $29 \gtrsim SNR_{\rm CFHTLS} \gtrsim 7$ (corresponding to $22.5
  < \zmag < 24$ for CFHTLS).
Therefore, \Chitah\ is able to detect \anguita, a faint small-separation quad, as a lens
candidate successfully.

\section{Summary and Discussions}
\label{sec:summary}

We have built a novel robot, \Chitah, to classify lensed candidates in
imaging surveys via modeling of the image configuration.
We use simulated CFHTLS-Wide-like lens systems from Space Warps 
to study the performance of \Chitah. 
The classification strategy is divided into four steps. 
First of all, we disentangle lens galaxy light and multiple quasar images using color information.
Secondly, we measure the lens center and the quasar image positions.
Thirdly, through the quasar image configuration, we separate the targets into two groups: potential quads and potential doubles.
Lastly, we model the potential quad/double image configuration via an
SIE/SIS lens distribution, and use the resulting $\chi^2$ from the
model to classify the lens.  We can choose an appropriate value for
the $\chith$ to separate lens and non-lens classifications (objects
with $\chi^2<\chith$ are classified as lenses whereas objects with
$\chi^2>\chith$ are classified as non-lenses). 

After testing \Chitah\ on simulated CFHTLS-Wide-like data we draw the following conclusions:
\begin{enumerate}
  \item  The optimal threashold of $\chith$ can be set to $\chith \sim 4$ for quad classification, 
and $\chith\sim1$ for double selection for imaging surveys with
  image qualities similar to that of CFHTLS.  
  \item \Chitah\ can hunt down much purer lens candidates for quads
    than doubles.  
  \item For bright quads with large image separations
    ($\rein>1\farcs1$) simulated with Gaussian PSFs, we achieve a high TPR ($\sim90\%$) and a low FPR
    ($\sim3\%$).  For the faint large-separation quads, \Chitah\ is also able to detect them very well with TPR$>$80\% and FPR$<$5\%. 
  \item We detect a sharp drop of TPR as $\rein$ becomes smaller than $\sim0\farcs5$
    (i.e., with quasar image separations $\lesssim1''$),
    which corresponds roughly to the PSF seeing of the mock lenses.
    The performance of \Chitah\ is thus set by the image quality.
  \item Relative to the Gaussian PSF, the extended wings of Moffat PSF decrease the TPRs by a few per cent.
  \item We feed the real lens system \anguita\ to \Chitah, and \Chitah\ successfully classifies it as a quad system.
\end{enumerate}

Having a fast \Chitah\ for lens classification is one of our goals
of the robot development.  Based on the simple yet robust scheme
(outlined in \sref{sec:procedure}), \Chitah\ takes about 5 s to
classify one object on an Intel i7 3.2GHz CPU.  This translated to
about a week to search through a million objects with an 8-core CPU.

To achieve a fast strategy, we simply employ the source plane
$\chi^2_{\rm src}$ in \eref{equ:chi2src}, 
 and add the lens light center as prior $\chi^2_{\rm c}$ in
 \eref{equ:chi2c}.  Despite the simplicity, we can detect lens candidates
with high TPR and low FPR based on our simulations.  A possible way to
enhance the algorithm even further would be to obtain $\chi^2_{\rm
  src}$ by considering the full magnification tensor \citep[e.g., see
Appendix 2 of][]{Oguri10}.  
Flux ratios are possible additional observational constraints for
modeling, although the flux ratio uncertainty would need to
accommodate the typical anomalous quasar flux ratios.  
Moreoever, the ellipticity and position angle of the lens light profile may also be added as priors, 
but the complexity may slow down the efficiency of the algorithm.
For exotic lenses that cannot be readily described by an SIE profile, 
one may also equip \Chitah\ with other lens models, or use additional bands (e.g., near infrared) or different band
combinations to identify better these rarer lensed objects.  Exotic
lenses are generally difficult to find with automated algorithms, and
are usually easier to spot via visual inspections, such as through
Space Warps.

Current surveys such as the HSC survey and DES are imaging a wide area of the
sky (thousands of square degrees) in multiple bands, and in the
future the LSST will image the
entire southern sky.  HSC, DES, and LSST have similar imaging
bands to the CFHTLS-Wide-like lenses considered in this paper, so
\Chitah\ is readily applicable to these current/future imaging surveys.  
There will be hundreds of new quasar lens
systems in these ongoing surveys and thousands of quasar lenses in
LSST \citep{OguriMarshall10}, increasing the existing sample by at
least two orders of magnitude.   
We expect \Chitah\ to be a good and
efficient hunter of new lenses in these surveys.


\section*{Acknowledgments}

We would like to thank Adriano Agnello, Claire Lackner, Peter
Schneider, and Tommaso Treu for useful
discussions.
We would also like to thank the anonymous referee for the constructive comments on this work.
We are grateful to Bau-Ching Hsieh, Robert Lupton, Naoki Yasuda, and the HSC software 
team for their help in accessing the Suprime-Cam images of \anguita.
We are also thankful to the Space Warps collaboration for sharing an early release of the duds sample.
J.H.H.C.~would like to thank Ui-Han Zhang for algorithm support. 
J.H.H.C.~and S.H.S.~acknowledge support from the Ministry of Science and Technology in Taiwan via grant MOST-103-2112-M-001-003-MY3.
T.C.~acknowledges the National Science Council of Taiwan via grant 100-2112-M-002-018-MY3.
A.M.~is supported by World Premier International Research Center Initiative (WPIInitiative), MEXT, Japan, 
and also acknowledges the support of the Japan Society for Promotion of Science (JSPS) fellowship.
The work of P.J.M.~ was supported in part by the U.S.~Department of
Energy under contract number DE-AC02-76SF00515. 
M.O.~acknowledges support in part by World Premier International
Research Center Initiative (WPI Initiative), MEXT, Japan, and
Grant-in-Aid for Scientific Research from the JSPS (26800093).
%


\bibliography{ChitahPaper.bib}
\bibliographystyle{apj}


\label{lastpage}
\end{document}